\documentclass[11pt]{article}
\usepackage[algoruled]{algorithm2e}
\usepackage{amsthm,amsmath,amssymb}
\usepackage[symbol]{footmisc}
\usepackage[margin=1in]{geometry}
\usepackage{graphicx}
\usepackage{natbib}
\usepackage[hidelinks,breaklinks,hypertexnames=False]{hyperref}
\usepackage{subfig}
\newtheorem{theorem}{Theorem}[]

\newtheorem{remark1}[theorem]{Remark}

\DeclareMathOperator{\E}{\mathop{}\mathbb{E}}

\renewcommand{\epsilon}{\varepsilon}
\renewcommand{\phi}{\varphi}
\newcommand{\negphantom}[1]{\settowidth{\dimen0}{#1}\hspace*{-\dimen0}}

\title{Measuring multi-calibration}
\author{Ido Guy, Daniel Haimovich, Fridolin Linder, Nastaran Okati,
        Lorenzo Perini, Niek Tax,\\and Mark Tygert}
\date{Central Applied Science at Meta
and Fundamental Artificial Intelligence Research at Meta}

\begin{document}

\maketitle

\begin{abstract}
A suitable scalar metric can help measure multi-calibration,
defined as follows. 
When the expected values of observed responses are equal
to corresponding predicted probabilities, the probabilistic predictions
are known as ``perfectly calibrated.''
When the predicted probabilities are perfectly calibrated simultaneously
across several subpopulations, the probabilistic predictions
are known as ``perfectly multi-calibrated.''
In practice, predicted probabilities are seldom perfectly multi-calibrated,
so a statistic measuring the distance from perfect multi-calibration
is informative. A recently proposed metric for calibration,
based on the classical Kuiper statistic, is a natural basis
for a new metric of multi-calibration and avoids well-known problems
of metrics based on binning or kernel density estimation.
The newly proposed metric weights the contributions of different subpopulations
in proportion to their signal-to-noise ratios; data analyses' ablations
demonstrate that the metric becomes noisy when omitting
the signal-to-noise ratios from the metric.
Numerical examples on benchmark data sets illustrate the new metric.
\end{abstract}

\section{Introduction}
\label{intro}

\subsection{Informal definitions of calibration and multi-calibration}

Predicted probabilities are known as ``perfectly calibrated''
when the expected values of corresponding observed responses
are equal to the predicted probabilities.
In perfectly calibrated predictions of weather, for example,
it would rain on precisely 80\% of the days predicted to have an 80\% chance
of rain (and similarly for any other percentage).
Good calibration helps quantify uncertainty and assign levels of confidence
to predictions. There are many metrics
for measuring calibration of a given set of observations;
see, for example, the work of~\cite{arrieta-ibarra-gujral-tannen-tygert-xu}.

Good calibration of the entire population of observations
need not imply good calibration in any particular subpopulation.
The task of ensuring good calibration for sundry specified subpopulations
simultaneously has been known as ``multi-calibration''
due to~\cite{hebert-johnson-kim-reingold-rothblum}.
The present paper proposes metrics for measuring multi-calibration.

The present paper builds and follows up
on theoretical advances of~\cite{haghtalab-jordan-zhao}.
The latter showed that accounting for statistical fluctuations is critical.
Omitting normalization for signal-to-noise ratios
(as in other extant works on multi-calibration) results in estimates
from finite samples that are mainly noise in the setting
of discrete classification; Subsection~\ref{discussion} illustrates this
in detail via numerical experiments. The present paper
proposes practical procedures for measuring multi-calibration,
complete with the weighting by signal-to-noise ratios
from the theorems of~\cite{haghtalab-jordan-zhao}.
For data sets containing only finitely many observations,
extensive earlier work reviewed in Appendix~\ref{ece_ici}
enables sample-efficient measurement of calibration,
which the present paper extends to multi-calibration.
The following subsection of this introduction reviews
key insights of~\cite{haghtalab-jordan-zhao}.

\subsection{Purposes of metrics for multi-calibration}

Any metric for multi-calibration merely complements without replacing
traditional metrics of performance.
This is key in constructing metrics for multi-calibration. Other metrics 
could include precision and recall, or sensitivity and specificity.
That said, the information and signals in metrics of multi-calibration
may help improve accuracy and other metrics,
even when multi-calibration is of little direct interest.
After all, multi-calibration inherently involves specification
of subpopulations or covariates of interest, thus incorporating a-priori,
domain-specific knowledge and insights.

Indeed, metrics of calibration for the full population alone
(without consideration for any proper subpopulation) can obscure signals
from different subpopulations that are suspected to impact accuracy;
suspecting such impact on accuracy is common
due to domain-specific knowledge.
Breaking the full population into subpopulations corresponding
to different geographical regions or different human attributes
often reveals information that looking at the full population may obscure
due to noise (noise can cause the different subpopulations
to interfere with each other's estimates).

So, any metric for multi-calibration ideally should account
for how the calibrations of the various subpopulations are likely
to contribute to the accuracy for the full population.
The metric proposed in the present paper takes this into account.
The metric weights the contribution from each subpopulation
under consideration in inverse proportion to the standard deviation
of the corresponding metric of calibration for that single subpopulation.
The standard deviation is calculated under the null hypothesis
of perfect calibration. This results in smaller subpopulations
contributing less, with the contribution to the overall metric being roughly
in proportion to the square root of the effective number of observations
in the subpopulation, as advocated earlier by~\cite{haghtalab-jordan-zhao}.
Other works omit this weighting that \cite{haghtalab-jordan-zhao}
proved to be essential.

To be precise, this noise level for estimates from a subpopulation is
inversely proportional to the square root of the number of observations
(or to the square root of the total weight of the observations
for weighted samples)
when drawing the predicted probabilities from a distribution
that is independent of the number of observations. The present paper details
how to handle any arbitrary distribution of the predicted probabilities,
whether or not the underlying distribution depends
on the number of observations.

\subsection{Maximal benefits in metrics for multi-calibration}

To combine estimates of calibration from many subpopulations,
the metric for multi-calibration proposed below takes a maximum
over all the given subpopulations.
Specifically, the metric first measures calibration for every one
of the specified subpopulations, then weights them
by their signal-to-noise ratios, and finally returns the maximum
over all subpopulations.
Users typically specify as many subpopulations as they can conceive.
Users are welcome to include as many potentially
causal connections as they can conceive, forming subpopulations according
to cuts along thresholds of covariates known or suspected a-priori
to be causally important.
With the normalization by the standard deviation discussed above,
taking a maximum focuses power on the largest subpopulations
when their calibration is poor, while reserving power
for the smaller subpopulations when the largest subpopulations
are already well-calibrated. Users generally address gross problems
with calibration before drilling down into more refined problems;
the metric proposed below measures progress made this way.
And interpreting a maximum is trivial, focusing attention
on the most problematic subpopulations.
Taking a maximum to focus on the worst-case has been popular with others,
such as~\cite{gohar-cheng}, \cite{hansen-devic-nakkiran-sharan},
and~\cite{shabat-cohen-mansour}.
This also aligns with~\cite{hebert-johnson-kim-reingold-rothblum}
and~\cite{haghtalab-jordan-zhao}, who consider the full population to be
well--multi-calibrated only when every single subpopulation considered
is well-calibrated.

\subsection{Benefits of good multi-calibration}

Many benefits accrue from enhancements to multi-calibration (that is,
from enhancing the calibration of each of many subpopulations simultaneously).
One benefit could be called ``in-distribution generalization.''
Indeed, such good multi-calibration enhances the weighted average accuracy
across the population when re-weighted according to different distributions.
With only good global calibration and average accuracy,
averaged over the full population, nothing guarantees good performance
across different weightings of the examples in the data set.
The ability to attain good performance without re-training
has various mathematical formulations, including the ``universal adaptability''
of~\cite{kim-kern-goldwasser-kreuter-reingold} and~\cite{ye-li},
as well as the ``out-of-distribution generalization''
of~\cite{wu-liu-cui-wu}.
Moreover, multi-calibration can automatically adjust for problematic outliers,
as part of the ``omniprediction'' of~\cite{gopalan-kim-reingold} and others.

\subsection{Further related work}

The work of~\cite{hebert-johnson-kim-reingold-rothblum} introduced
the term, ``multi-calibration,'' along with an algorithm
for improving multi-calibration that leveraged boosting.
A software package, ``{\tt mcboost},'' based on that algorithm was developed
by~\cite{pfisterer-kern-dandl-sun-kim-bischl}.
A related notion is ``multi--accuracy-in-expectation,''
with the relation developed by~\cite{kim-ghorbani-zou} and others.
Multi--accuracy-in-expectation concerns the average accuracy
for each of several subpopulations, while multi-calibration concerns
calibration for the subpopulations.
The present paper focuses on metrics of multi-calibration
rather than algorithms for improving multi-calibration.

Appendix~\ref{ece_ici} reviews popular yet irreparably problematic metrics
for calibration, namely the ICI (``integrated calibration index'')
of~\cite{austin-steyerberg} and the ECE (``empirical calibration error,''
``estimated calibration error,'' or ``expected calibration error'')
and reliability diagrams reviewed by~\cite{brocker-smith}, \cite{brocker},
\cite{wilks}, and others.
As reviewed in Appendix~\ref{ece_ici}, the ECE and ICI are inherently limited
for the case of assessing calibration from a finite sample
of an infinite data set,
so would also be inappropriate for assessing multi-calibration.
Instead, the present paper uses an approach due
to~\cite{kolmogorov} and~\cite{smirnov} and to~\cite{kuiper},
later extended to calibration by~\cite{arrieta-ibarra-gujral-tannen-tygert-xu},
\cite{tygert_full}, and others.
The Kuiper metric avoids making an explicit trade-off
between averaging away random noise and the statistical power
to resolve variations as a function of the predicted probabilities.
\cite{lee-huang-hassani-dobriban} proved rigorously that
such an undesirable trade-off is unavoidable in methods such as the ECE or ICI
which leverage binning into histograms or kernel density estimation.

\subsection{The presentation in the present paper}

This paper adheres to the terminological convention of machine learning
that ``classification'' refers to predictions into discrete classes,
while ``regression'' refers to predictions that can take
on a continuum of real values. In contrast, the convention among some
statisticians is that ``regression'' refers to both classification
into discrete classes as well as predictions that can take on a continuum
of values. ``Logistic regression'' actually refers to classification;
the terminology, ``logistic regression,'' comes from statisticians.
For simplicity and concreteness, the present paper focuses
on classification into the two classes of Bernoulli variates,
0 (``failure'') and 1 (``success''), mentioning
the straightforward generalization to general regressions
briefly in Appendix~\ref{estimates}.

The remainder of the present paper has the following structure:
Section~\ref{methods} introduces the main methodology and metric.
Section~\ref{results} presents results from applying the methods
to a couple benchmark data sets.\footnote{An open-source software package
that can automatically reproduce all results reported below is available at
\url{https://github.com/facebookresearch/McMetric}}
Section~\ref{conclusion} summarizes the findings.
Appendix~\ref{proportional} points to the related notion
of ``proportional'' multi-calibration
introduced by~\cite{la-cava-lett-wan}.
Appendix~\ref{low_prevalence} proposes an adjustment suitable
for use in the case of significant imbalance between the numbers
of observed instances of the different classes.
Appendix~\ref{estimates} mentions a possible generalization
to observed responses that can take real values other than the 0 (failure)
and 1 (success) from Bernoulli trials.
Appendix~\ref{ece_ici} reviews the ECE and ICI and their intrinsic limitations.
Appendix~\ref{boosting} reviews schemes for improving multi-calibration
based on boosting or augmentation.
Appendix~\ref{synthetic} constructs a synthetic data set
relevant for unit testing of the associated software package.
Appendix~\ref{additional} supplements Section~\ref{results}
with tables of further numerical results.

\section{Methods}
\label{methods}

This section introduces the methodology of the present paper.
Subsection~\ref{notation} sets notation used throughout.
Subsection~\ref{mainmetric} introduces the main metric for multi-calibration.
Subsection~\ref{gen} details an algorithm for generating subpopulations
automatically based on given covariates.

\subsection{Notation}
\label{notation}

This subsection sets notational conventions used throughout the present paper.
All results mentioned in the present subsection are discussed, for example,
by~\cite{tygert_full}.

Consider predicted probabilities $S_1 \le S_2 \le \dots \le S_{n_0}$,
known as ``scores,'' and corresponding numbers
$R_1$, $R_2$, \dots, $R_{n_0}$, known as ``responses,''
each of which can take either the value $0$ or the value $1$.
Consider also corresponding positive real numbers
$W_1$, $W_2$, \dots, $W_{n_0}$, known as ``weights'' for a weighted sampling.
(If the sampling is unweighted, then the weights are uniform,
with $W_1 = W_2 = \dots = W_{n_0}$.)
The scores will be viewed as deterministic, not random,
while the responses will be viewed as random variables that are independent.
Consider $\ell$ subpopulations specified by indices
$i_1^k < i_2^k < \dots < i_{n_k}^k$ for $k = 0$, $1$, $2$, \dots, $\ell$.
``Subpopulation'' 0 will be the full population,
so $i_1^0 = 1$, and $i_2^0 = 2$, \dots, and $i_{n_0}^0 = n_0$.
The fact that every subpopulation is part of the full population yields that
$n_k \le n_0$ for $k = 0$, $1$, \dots, $\ell$.
Requiring $n_k$ to be greater than, say, 9,
for all $k = 0$, $1$, \dots, $\ell$, can help ensure that statistics estimated
for the subpopulations are not mostly random noise, too.

Consider any $k = 0$, $1$, \dots, $\ell$.
The cumulative differences for subpopulation $k$ are
\begin{equation}
C_j^k = \frac{\sum_{m=1}^j \left( R_{i_m^k} - S_{i_m^k} \right)
                           \cdot W_{i_m^k}}
             {\sum_{m=1}^{n_k} W_{i_m^k}}
\end{equation}
for $j = 1$, $2$, \dots, $n_k$; obviously, $-1 \le C_j^k \le 1$
for $j = 1$, $2$, \dots, $n_k$. We also set $C_0^k = 0$.
The Kuiper metric for subpopulation $k$ is
\begin{equation}
\label{kuiper1}
D_k = \max_{0 \le j \le n_k} C_j^k - \min_{0 \le j \le n_k} C_j^k;
\end{equation}
clearly, $0 \le D_k \le 1$, providing a universally normalized measure
of the effect size for miscalibration.
{\it The Kuiper metric is also equal to the absolute value
of the total miscalibration, totaled over the interval of scores
for which the absolute value of the total is greatest.} That is,
\begin{equation}
\label{kuiper2}
D_k = \max_{1 \le p \le q \le n_k}
\left| \frac{\sum_{m=p}^q \left( R_{i_m^k} - S_{i_m^k} \right)
                          \cdot W_{i_m^k}}
            {\sum_{m=1}^{n_k} W_{i_m^k}} \right|.
\end{equation}

\subsection{Metric for multi-calibration}
\label{mainmetric}

This subsection introduces the main metric for multi-calibration
that the present paper proposes. The appendices discuss possible alternatives
modifying the basic proposition of the present subsection
that could be relevant for special circumstances.

The metric detailed in the present subsection measures calibration
for every one of the specified subpopulations, then normalizes them
by their signal-to-noise ratios, and finally returns the worst case,
that is, the maximum over all subpopulations. Focusing on the worst case
identifies which subpopulation has the greatest potential for improvement,
providing a readily interpretable and actionable diagnostic.

Thus, the multi-calibration metric which considers all subpopulations
on an equal footing, in proportion to their signal-to-noise ratios, is
\begin{equation}
\label{multimetric}
M = \max_{0 \le k \le \ell} \frac{D_k \cdot \sigma_0}{\sigma_k},
\end{equation}
where
\begin{equation}
\label{stddev}
\sigma_k = \frac{\sqrt{\sum_{j=1}^{n_k} \left[S_{i_j^k}\left(1-S_{i_j^k}\right)
                                        \left(W_{i_j^k}\right)^2\right]}}
           {\sum_{j=1}^{n_k} W_{i_j^k}}
\end{equation}
for $k = 0$, $1$, \dots, $\ell$.
Indeed, under the null hypothesis that $R_j$ is drawn independently
from the Bernoulli distribution whose mean is $S_j$,
for all $j = 1$, $2$, \dots, $n_0$, the standardized value
of $D_k$ is proportional to the factor $(D_k/\sigma_k)$ in~(\ref{multimetric})
that does not involve the full population $k = 0$.
This normalization of $D_k$ ensures that its distribution
under the null hypothesis is similar to the distribution
of the absolute value of a standard normal variate.
Under the null hypothesis, after all,
$C_0^k$, $C_1^k$, $C_2^k$, \dots, $C_{n_k}^k$
is a driftless random walk and the standard deviation of $C_{n_k}^k$
is $\sigma_k$ defined in~(\ref{stddev}).
When interpreting~(\ref{multimetric}) and~(\ref{stddev}),
remember that the scores and weights are non-random while the responses
are independent random variables.

The expected value of $D_k$ calculated under the null hypothesis
is at most $2 \sqrt{2 / \pi} \approx 1.6$ times the quantity in~(\ref{stddev})
and rapidly converges to that upper bound in the appropriate asymptotic limits,
as proven in Section~3.1 of~\cite{tygert_pvals}.
The dependence in~(\ref{multimetric}) on the expression in~(\ref{stddev})
incorporates the lessons of~\cite{haghtalab-jordan-zhao}.

The factor $\sigma_0$ in~(\ref{multimetric})
that does involve the full population $k = 0$
ensures that the weight of $D_0$ in the maximum from~(\ref{multimetric}) is 1
(that is, the argument to the max for $k = 0$ is simply $D_0$).
The most direct measure of effect size for the full population is $D_0$,
and so the metric $M$ in~(\ref{multimetric}) directly assesses effect size
with reference to the same universal scale ranging from 0 to 1
as for any Kuiper metric.

When the denominator in~(\ref{multimetric}) is 0 or nearly 0
and the numerator is also 0 or nearly 0, the resulting quotient
should be taken to be 0.
This consideration is critical for implementation
in finite-precision arithmetic (such as floating-point or fixed-precision).

\subsection{Algorithm for automatically generating subpopulations}
\label{gen}

This subsection proposes a randomized method for generating subpopulations
which correspond to various ranges of the values of covariates.
By construction, the subpopulations generated reflect information contained
in the given covariates. This allows for the automatic generation
of subpopulations given only variables that incorporate information
of domain- or application-specific interest.

The algorithm aims to sample at random paths in a binary tree.
The high-level strategy is to split recursively according to values
of the covariates, building a binary tree for which each subpopulation
generated corresponds to a path starting from the root.
The generation includes all such subpopulations along the path from the root
to a leaf; that is to say, the subpopulations generated include one
for every subpath starting at the root.

The root of the tree corresponds to the full population.
We then choose one of the covariates at random and retain only those members
whose values of the covariate are less than the median of their unique values
or instead are greater than or equal to the median (with the choice
between less than and greater than or equal to the median made at random).
We split subpopulations recursively, always choosing the next covariate
at random (with replacement). The median of the ``unique values''
takes the median of the list of distinct values of the covariate
for the members of the subpopulation prior to splitting.
Again, all subpopulations along a single path from the root to a leaf
are retained for the final set generated; the leaf corresponds
to a subpopulation whose number of members is no less
than a user-specified threshold (typically 10 or so,
though the threshold can be as low as 1, if the user prefers).

For ordinal covariates, namely, variates for which their values
come with a total ordering, the splits of the previous paragraph
are uniquely, unambiguously defined. Ordinal variates include
real-valued variates, integer-valued variates, and more generally
any categorical variates for which the categories are ordered.
For nominal variates, which are categorical variates for which the categories
have no natural ordering, we re-order the categories at random
every time we start building a path from the root to a leaf.
We keep the same (randomized) order of the categories for each nominal variate
until we reach a leaf, then re-randomize the orders for all nominal variates
when starting a new path back at the root.

For implementations in practice, further optimizations are useful.
There is no need to save in memory all subpopulations generated;
instead, each subpopulation can be generated without saving any of the others.
Furthermore, recording the splits used for subpopulations generated earlier
can enable efficiently checking whether the same splits were used before
(in which case generating the same subpopulation again would be redundant).
Recording only the splits will typically require far less memory
than storing all the indices in the subpopulations;
after all, there can be only just so many levels in a binary tree.

Algorithm~\ref{generation} provides pseudocode that omits the refinements
mentioned in the previous paragraph.
The Python codes in the associated GitHub repository
include all the refinements.\footnote{Open-source Python scripts and libraries
that can reproduce all results of the present paper are available at
\url{https://github.com/facebookresearch/McMetric}}
The refinements are for efficiency only,
not required for appreciation of the basic idea.

\LinesNumbered
\begin{algorithm}
\caption{Randomized generation of subpopulations}
\label{generation}
\DontPrintSemicolon
\KwIn{A positive integer $m$ that will be the minimum possible number
of members in a subpopulation, a positive integer $\ell$
that will be the number of subpopulations generated, and $p$ sets
of real numbers $a_1^{(1)}$, $a_2^{(1)}$, \dots, $a_{n_0}^{(1)}$;\,
$a_1^{(2)}$, $a_2^{(2)}$, \dots, $a_{n_0}^{(2)}$;\,\, \dots;
$a_1^{(p)}$, $a_2^{(p)}$, \dots, $a_{n_0}^{(p)}$
that are the values of ``covariates'' (also known as ``features'')
}
\KwOut{A sequence of subpopulations $I_0$, $I_1$, $I_2$, \dots, $I_{\ell}$,
where each subpopulation refers to a set of indices
specifying the subpopulation, namely,
$I_k$ is the set of indices $i_1^k$, $i_2^k$, \dots, $i_{n_k}^k$}

Set $k = 0$.

\Repeat{$k \ge \ell$.}{
For every nominal covariate, say the $j$th covariate,
re-order at random the values taken by the full population's
$a_1^{(j)}$, $a_2^{(j)}$, \dots, $a_{n_0}^{(j)}$.
(Maintain the original order of each ordinal covariate;
randomize only the ordering of the categories for every nominal covariate.)

Set $I_k$ to be the set of all indices, 1, 2, \dots, $n_0$
(recall that $i_1^0 = 1$, and $i_2^0 = 2$, \dots, and $i_{n_0}^0 = n_0$).

\Repeat{there are strictly fewer than $m$ members in $I_k$.}{
Choose an integer $j$ uniformly at random from $1$, $2$, \dots, $p$,
thus choosing one of the covariates at random.

Calculate the median $v$ of the distinct values that the values
$a_{i_1^k}^{(j)}$, $a_{i_2^k}^{(j)}$, \dots, $a_{i_{n_k}^k}^{(j)}$
of the $j$th covariate take.

With probability $1/2$, set $I_{k+1}$ to be the members of $I_k$
whose corresponding values
$a_{i_1^k}^{(j)}$, $a_{i_2^k}^{(j)}$, \dots, $a_{i_{n_k}^k}^{(j)}$
are less than the median $v$ of their distinct values;
while for the other probability $1/2$, set $I_{k+1}$ to be the members
of $I_k$ whose covariate values are greater than or equal to the median $v$.

Increment $k$ by 1 (that is, add 1 to $k$).
}

Decrement $k$ by 1 (that is, subtract 1 from $k$).
}

\Return{the subpopulations $I_0$, $I_1$, $I_2$, \dots, $I_{\ell}$.}

\end{algorithm}

\section{Results and discussion}
\label{results}

This section presents and discusses the results of experiments
on two standard benchmark data sets, namely the 1998 KDD Cup
and the 2019 American Community Survey
of the United States Census Bureau.\footnote{Open-source software
that can automatically reproduce all results reported in the present paper
is available at
\url{https://github.com/facebookresearch/McMetric}}
Subsection~\ref{set-up} details the configurations of the experiments.
Subsection~\ref{models} describes the machine-learned models considered.
Subsection~\ref{acs2019} describes the 2019 American Community Survey;
Subsection~\ref{kddcup1998} describes the 1998 KDD Cup.
Subsection~\ref{discussion} discusses the results.
Further examples are available in Appendices~\ref{synthetic}
and~\ref{additional}.

The purpose of the present section is to illustrate the methods
of the previous section, Section~\ref{methods}.
In particular, Subsection~\ref{discussion} demonstrates the importance
of normalizing by signal-to-noise ratios in the metrics for multi-calibration;
omitting such normalization results in metrics that are mostly noise.
The extensive tables of Appendix~\ref{additional} refer to the metrics
which omit such normalization as ``multi-ablate,''
as appropriate for an ablation to demonstrate the necessity of normalization.
The present section also discusses the results of calibrating
via different methods and finds that isotonic regression is a strong baseline;
Appendix~\ref{additional} considers various methods designed specifically
to improve multi-calibration, too.
Finally, in most cases the metric for multi-calibration
of~(\ref{multimetric}) is significantly greater than the metric for calibration
of~(\ref{kuiper1}) and~(\ref{kuiper2}),
demonstrating that the metric for multi-calibration discovers miscalibration
beyond that measured in the metric for calibration of the full population.

Appendix~\ref{additional} includes a comprehensive battery of tests.
The appendix considers all combinations of three base machine-learned models
coupled with five different methods for calibration
(plus the original base with no explicit calibration)
coupled with the two data sets considered (with three different responses
considered for the first data set)
coupled with metrics of calibration and multi-calibration
(both with and without weighting by signal-to-noise ratios).
All together, this yields $3 \cdot 6 \cdot 4 = 72$ examples
across every possible configuration, with each example evaluated
across all metrics. Subsection~\ref{discussion} summarizes conclusions
drawn from the results of the experiments.

\subsection{Experimental set-up}
\label{set-up}

This subsection provides details about the experiments conducted.
The following subsection details the machine-learned models used.

All results reported are on a test set obtained by reserving
a quarter of the data examples for the test set.
We set the parameter $\ell$ in~(\ref{multimetric})
and the implementation of Algorithm~\ref{generation}
($\ell$ is the maximum number of subpopulations considered)
to be $\ell =$ 1,000.
We set the parameter $m$ in Algorithm~\ref{generation}
($m$ is the minimum size of a subpopulation) to be $m = 10$.

The cross-validations for Platt scaling and for isotonic regression
leverage the Python package {\tt scikit-learn} of~\cite{scikit-learn}, using
{\tt sklearn.calibration.CalibratedClassifierCV} with the default settings.
The default settings perform five-fold cross-validation, holding out a fifth
of the original training examples as a validation set
for each of the five folds.
The Platt scaling and isotonic regression use the validation set
to calibrate the model that is fitted on the other four fifths
of the original training set.
The final calibrated estimator is the ensemble average
over the five pairs of models and calibrations fit via the five folds.

\subsection{Statistical models}
\label{models}

This subsection details the machine-learned models considered
in the numerical experiments of the present section.
All values of metrics reported pertain to scores that are predictions made
with these models.

We consider three models for classification:
logistic regressions, decision trees
(also known as ``classification and regression trees'' --- CART),
and the gradient-boosted decision-trees of LightGBM
of~\cite{ke-meng-finley-wang-chen-ma-ye-liu}.
For the first two, we use the implementations in the Python package
{\tt sklearn} of~\cite{scikit-learn}; for the last,
we use the implementation in the Python package {\tt lightgbm}
of~\cite{ke-meng-finley-wang-chen-ma-ye-liu}.

We stick with the default parameters, aside from the following choices
(which our experiments indicate are close to optimal
for the data sets considered):

\begin{itemize}
\item
{\tt sklearn.linear\_model.LogisticRegression}:
{\tt tol} = 0.01,\ {\tt solver} = saga,\ {\tt max\_iter} = 1000;
\item
{\tt sklearn.tree.DecisionTreeClassifier}:
{\tt max\_depth} = 4;
\item
{\tt lightgbm.LGBMClassifier}:
{\tt max\_depth} = 3,\ {\tt num\_leaves} = $2^{\tt max\_depth}$,\
{\tt n\_estimators} = 10.
\end{itemize}

In general, none of the models was dramatically superior to the others
in the empirical results reported by the tables. Decision trees
or their boosted versions in LightGBM look to be safe choices.

\subsection{American Community Survey of the U.S.\ Census Bureau}
\label{acs2019}

This subsection describes the first data set considered.
Appendix~\ref{additional} reports further numerical results from this data, via
Tables~\ref{acsTABLETLGBMClassifier}--\ref{acsSMARTPHONELogisticRegression}.
The present subsection includes two brief case studies for readers
who prefer not to dive into the comprehensive battery of experiments
in the appendix and its discussion in Subsection~\ref{discussion}.

The 2019 American Community Survey of the U.S.\ Census Bureau
samples households from across California; we focus on California's
Orange County alone except in the last three paragraphs
of the present subsection.\footnote{The microdata
of the American Community Survey is available
from the United States Census Bureau
at \url{https://www.census.gov/programs-surveys/acs/microdata.html}}
The sampling is weighted, so we retain only those households
whose weights (``WGTP'' in the microdata), personal income (``HINCP''),
adjustment factor to income (``ADJINC''), and number of people (``NP'')
are strictly positive (eliminating vacant addresses and group quarters),
leaving 10,680 households in the sample of Orange County.
We included 233 covariates from the data set. We normalized every covariate
by the maximum value that the retained households attain.
We consider predicting one of three response variates:
the presence of a tablet computer, the presence of a laptop computer,
and the presence of a smartphone in the household
(the response takes the value 1 when the associated computational device
is present in the household and takes the value 0 otherwise).
The covariates are also the features in Algorithm~\ref{generation},
so the parameter $p$ of Algorithm~\ref{generation} is $p = 233$
when analyzing Orange County.

With regard to all the different methods
for calibrating the predicted probabilities
considered in Appendix~\ref{additional},
the empirical results reported in the tables of Appendix~\ref{additional}
indicate that isotonic regression is often the best or nearly the best
among the methods considered. So the remainder of the present subsection
focuses on isotonic regression when performing calibration.

For a brief case study, we train a LightGBM model and then calibrate
via isotonic regression with cross-validation, when the response variate
is the presence of a tablet computer in the household.
Tables~\ref{acsTABLETLGBMClassifier}--\ref{acsTABLETLogisticRegression}
show that this configuration coupling LightGBM with isotonic regression
outperforms all other configurations.
In this case, the Kuiper metric prior to calibration is 0.04085,
while its expected value calculated under the null hypothesis
of perfect calibration is about 0.01609, for a ratio of 2.539.
The Kuiper metric after calibration is 0.01379,
while its expected value under the null hypothesis is about 0.01442,
for a ratio of 0.9562. Thus, isotonic regression greatly reduces
both the absolute value of the Kuiper metric as well as its ratio
relative to its expected value.

Similarly, an evaluation of the metric $M$ for multi-calibration
yields 0.07251 prior to calibration;
dividing by the standard deviation from~(\ref{stddev}) for the full population
(calculated under the null hypothesis) yields 7.192.
This metric $M$ of multi-calibration is 0.04085 after calibration;
dividing by the standard deviation yields 4.520.
Again, the isotonic regression drastically reduces both the absolute value
as well as the ratio relative to the expected value under the null hypothesis.
The argmax (worst-case) subpopulation prior to calibration arises
from splitting along the variables for the adjusted household personal income
and for one of the independent samples of the weights.
The argmax subpopulation after calibration involves several more levels
of refinement, splitting along the variables for the index in an ordered list
of the language spoken, for the sexes of the householders (ordered to cover
all cases), for an indicator that takes the value 1 when having a telephone
is imputed and 0 otherwise, and for two independent samples of the weights.
(The weights come with independent samples that help characterize
the sampling design --- the survey is not a full census but instead
arises from random sampling whose design the independent samples
of the weights help describe.)
Thus, the isotonic regression results in the argmax subpopulation being
much further down the binary tree of splits along covariates.

For completeness, we consider also all counties of California
(rather than only Orange County), together with an extra nominal covariate
that specifies the county for each individual household.
The larger data set which includes all of California
has a total of 134,094 households in its weighted sample.
For this larger data set, we use ten other covariates (so eleven in all),
instead of the hundreds considered earlier, in order to illustrate
both settings. The full list of covariates is the county
in which the members of the household reside, the number of people
in the household, the number of related children in the household,
the number of vehicles in the household, the number of rooms for the household,
as well as binary indicators of whether the household got food stamps,
has broadband access to the Internet, has high-speed access to the Internet,
has satellite access to the Internet, has a laptop computer,
and has a smartphone. We again consider predicting whether the household
has a tablet computer. As in the earlier case study,
we use LightGBM as the model and calibrate via isotonic regression
with cross-validation (isotonic regression again outperformed
across all metrics both Platt scaling and the three iterations of augmentation
from Appendix~\ref{boosting} with any of the three models
from Subsection~\ref{models}).

For this larger example, the test error without including any covariates at all
is 0.3471, the test error after training the model but before calibration
reduces to 0.2485, and the test error reduces still further to 0.2434
after calibration with isotonic regression.
The Kuiper metric is 0.05665 before calibration and 0.003738 after calibration
with isotonic regression. The corresponding expected values calculated
under the null hypothesis of perfect calibration ---
about 1.6 times the standard deviation from~(\ref{stddev}) ---
are 0.004970 and 0.004530. The corresponding numbers for an evaluation
of the metric $M$ of~(\ref{multimetric}) for multi-calibration are $0.05665$
and $0.01738$. Thus, calibration via isotonic regression with cross-validation
improves performance significantly across all metrics.

We could also consider ignoring the signal-to-noise ratio used for $M$
in~(\ref{multimetric}), reporting simply $\max_{0 \le k \le \ell} D_k$,
where $D_k$ is the Kuiper metric of~(\ref{kuiper1}) and~(\ref{kuiper2})
for subpopulation $k$.
When ignoring the signal-to-noise ratio, the metric becomes $.4069$
prior to calibration and $.5347$ after calibration,
while the expected values under the null hypothesis of perfect calibration
for the argmax subpopulation would be $0.2041$ and $.1862$
--- around half the observed values.
Given that 95\% confidence intervals extend roughly two standard deviations
with respect to the mean values, this shows that ignoring
the signal-to-noise ratio results in metrics that are very noisy.

\subsection{KDD Cup 1998}
\label{kddcup1998}

This subsection describes the other data considered.
Tables~\ref{kddcupLGBMClassifier}--\ref{kddcupLogisticRegression}
of Appendix~\ref{additional} report numerical results on the data described
in the present subsection.
The following subsection, Subsection~\ref{discussion}, discusses the results.

A national veterans organization in the U.S. conducted an experiment in 1994,
mailing solicitations for donations.
The 1998 Knowledge Discovery and Data-mining (KDD) Cup competition
leveraged this data.\footnote{The 1998 KDD Cup's data is available
at \url{https://kdd.ics.uci.edu/databases/kddcup98/kddcup98.html}}
We included 324 covariates from the data set for the 80,796 individuals
who got mailings in both 1995 and 1996 and donated at least once.
We used these covariates to predict whether the organization
deems the recipient to be a ``star donor''
(the data set conveniently provides a response variate that takes the value 1
when an individual is a ``star donor'' and takes the value 0 otherwise).
We normalized every covariate to range from 0 to 1,
first subtracting off the minimum value of the covariate.
The covariates are also the features in Algorithm~\ref{generation},
so the parameter $p$ of Algorithm~\ref{generation} is $p = 324$
in the present subsection.

As discussed in the following subsection, the empirical results reported
in the tables of Appendix~\ref{additional} demonstrate the importance
of weighting by signal-to-noise ratios in the metric for multi-calibration.
With regard to all the different methods
for calibrating the predicted probabilities
considered in Appendix~\ref{additional},
the tables of Appendix~\ref{additional} indicate
that the thrice-augmented decision-trees are often the best
or nearly the best among the methods considered.
However, calibration via isotonic regression is competitive.

\subsection{Discussion}
\label{discussion}

This subsection discusses the results and implications for methodology.

Isotonic regression for calibration appears to be competitive
in the numerical results reported by the tables for the data sets
considered here. Thus, isotonic regression can be a good baseline
for calibration, at the very least. Subsection~\ref{acs2019} above considered
this baseline in detail.
In general, combining different methods appears to yield
greater benefits than, say, using for the augmentations
of Appendix~\ref{boosting} the same statistical model
as used for training prior to any calibration.

Neglecting the signal-to-noise ratio when forming the metric
of multi-calibration yields the ``multi-ablate'' metric reported in the tables
of Appendix~\ref{additional},
which is also the simple maximum ($\max_{0 \le k \le \ell} D_k$)
whose discussion concludes Subsection~\ref{acs2019} above.
This metric is the same as the multi-calibration metric $M$
defined in~(\ref{multimetric})
except without any adjustment for the standard deviation in~(\ref{stddev}).
This ``multi-ablate'' metric looks to be completely useless;
its value is very large uniformly across all experiments and configurations.
Omitting adjustment for the signal-to-noise ratio in~(\ref{stddev})
causes the metric to focus on noise from the smallest subpopulations.

In fact, the values of the ``multi-ablate'' metric are seldom more than
a few times greater than the corresponding ``expectation'' values calculated
under the null hypothesis of perfect calibration.
This illustrates that neglecting the signal-to-noise ratio results
in noisy metrics, given that 95\% confidence intervals extend
about two standard deviations with respect to the expected values.
The conclusion of Subsection~\ref{acs2019} discussed this;
comparing the last two rows in each of Appendix~\ref{additional}'s
Tables~\ref{acsTABLETLGBMClassifier}--\ref{kddcupLogisticRegression}
gives 72 more examples.
This strongly confirms the lessons of~\cite{haghtalab-jordan-zhao}.

The metric $M$ defined in~(\ref{multimetric}) likely is most informative
when the subpopulations are calibrated reasonably well, that is,
when they are close to satisfying the null hypothesis of perfect calibration.
Weighting subpopulations in inverse proportion to the standard deviations
(under the null hypothesis) of their individual Kuiper metrics is most relevant
when the subpopulations do not deviate too significantly
from the null hypothesis.
The metric $M$ defined in~(\ref{multimetric}) effects this weighting and so
could be viewed as a regularized measure of calibration
for the full population, regularized with consideration for the expected levels
of noise on the metrics for the individual subpopulations.

When subpopulations are far from perfect calibration,
the multiplicative weighting by the reciprocal of the expected level of noise
in the metric $M$ will tend to suppress the values of the Kuiper metric
for smaller subpopulations that would otherwise have contributed strongly
to the maximum over subpopulations that $M$ reports.
Thus, $M$ focuses first on calibration for the full population
and would pay more attention to miscalibration of the smaller subpopulations
only as calibration improves --- when calibration of the full population
is poor, the maximum defining $M$ will tend to report that miscalibration
rather than any of the smaller subpopulations'.

Many users are likely to include subpopulations so small that noise becomes
a real concern, requiring some sort of weighting according
to statistical significance. Indeed, users may try to squeeze
the strongest possible statistical inferences out of the limited data
that they have. The original proposition
of~\cite{hebert-johnson-kim-reingold-rothblum} addresses the resulting issues
of statistical estimation by considering circuit complexity and oracles
(as in the computational complexity theory of theoretical computer science).
The present paper addresses statistical issues arising from statistics,
instead. The statistical issues come from the responses being discrete
when the task is classification (whereas the responses need not be discrete
when the task is regression more generally). Even when the responses
are discrete in a binary classification, the underlying probabilities
need not be always either 0 or 1. The statistical issues
arise from the noise inherent in the observed responses being only 0 or 1
while their expected values can take values strictly between 0 and 1, too.

\section{Conclusion}
\label{conclusion}

A natural metric for measuring multi-calibration across a set of subpopulations
is the maximum (over the set of subpopulations)
of the subpopulation's Kuiper metric,
each normalized by its standard deviation calculated under the null hypothesis
of perfect calibration. The Kuiper metric here refers to that measuring
miscalibration of the predicted probabilities
for the corresponding subpopulation.
There is also a straightforward algorithm for constructing subpopulations
based on the values of covariates
(``covariates'' are also known as ``features''),
detailed in Subsection~\ref{gen} and Algorithm~\ref{generation}.
Taking the maximum in~(\ref{multimetric}) and also multiplying
by the signal-to-noise ratio (the reciprocal of the standard deviation)
has many advantages,
discussed in Section~\ref{intro} and in Subsection~\ref{discussion};
the metric proposed above secures these advantages.
The following appendices suggest straightforward extensions required
to address so-called ``proportional multi-calibration,''
to address low-prevalence settings, and to address continuous regressions
rather than only discrete classifications.
The next appendix discusses advantages of the Kuiper metric
over the ECE and ICI.
The appendix after that describes a scheme for improving multi-calibration
based on boosting or augmentation.
The penultimate appendix provides a synthetic example with closed-form analytic
expressions for the metrics.
The final appendix provides further numerical illustrations on real data,
complementing those in Section~\ref{results} above.

\section*{Acknowledgements}

We would like to thank L\'eon Bottou, Kamalika Chaudhuri, and Adina Williams.

\appendix

\section*{Appendices}

Appendix~\ref{proportional} adapts the methodology of the present paper
to what is known as ``proportional'' multi-calibration.
Appendix~\ref{low_prevalence} modifies the methods for cases
of severe imbalance between the two classes (classes 0 and 1).
Appendix~\ref{estimates} generalizes the methods from discrete classifications
to continuous regressions.
Appendix~\ref{ece_ici} reviews problems with popular metrics
(the ECEs and ICIs) for calibration.
Appendix~\ref{boosting} reviews procedures for improving multi-calibration
based on boosting or augmentation.
Appendix~\ref{synthetic} calculates closed-form analytic expressions
of the metrics for a synthetic data set, appropriate for unit testing
of the software.
Appendix~\ref{additional} supplements the results reported
in Subsection~\ref{acs2019} with tables of further results.

\section{Proportional multi-calibration}
\label{proportional}

``Proportional'' calibration and its (entirely straightforward) generalization
to multi-calibration was proposed by~\cite{la-cava-lett-wan}.
Proportional calibration is the special case of weighted calibration
in which the weight is inversely proportional to the score, that is,
$W_j = 1/S_j$, where $W_j$ is the weight and $S_j$ is the score.
A regularized variant was also proposed by~\cite{la-cava-lett-wan};
the variant introduces a regularization parameter
--- a positive real number $\rho$ --- and corresponds
to the weighting $W_j = 1/\rho$ (when $0 \le S_j \le \rho$)
and $W_j = 1/S_j$ (when $S_j > \rho$). Another way to regularize that is
very similar is to take $W_j = 1 / (S_j + \rho)$.
This weighting highlights false positives more than false negatives,
to the extent that the scores are predicting the classes correctly.
The following appendix suggests another method for emphasizing false positives
when the prevalence is low for the class for which responses take the value 1.

\section{Low prevalence}
\label{low_prevalence}

For simplicity, consider the case in which the original data is unweighted.
The observed prevalence of the class for which responses take the value 1 is
then simply the average response $A = \sum_{j=1}^{n_0} R_j / n_0$.
Emphasizing false positives more than false negatives may be desirable
when the average $A$ is low.
If the responses are assumed to be correct,
then emphasizing false positives is possible by introducing weights,
$W_j = 1/A$ for $j$ such that $R_j = 1$, and $W_j = 1$ for $j$
such that $R_j = 0$.

\section{Variance estimates for regressions}
\label{estimates}

When the responses $R_1$, $R_2$, \dots, $R_{n_0}$ can take real values
other than only 0 and 1, the standard deviation of $C_{n_k}^k$
under the null hypothesis that $\E[R_j] = S_j$ for $j=1$, $2$, \dots, $n_0$,
is no longer the value from~(\ref{stddev}) but instead a good estimate becomes
\begin{equation}
\label{estimator}
\sqrt{\frac{\sum_{j=1}^{n_k - 1}
      \left( R_{i_j^k} - S_{i_j^k} - R_{i_{j+1}^k} + S_{i_{j+1}^k} \right)^2
      \left( W_{i_j^k} + W_{i_{j+1}^k} \right)^2}
     {4 \left( \sum_{j=1}^{n_k} W_{i_j^k} \right)
        \left( W_{i_1^k} + W_{i_{n_k}^k}
             + 2 \sum_{j=2}^{n_k - 1} W_{i_j^k} \right)}}
\end{equation}
for $k = 0$, $1$, \dots, $\ell$.
The factor of 4 in the denominator of~(\ref{estimator})
compensates for including in the numerator the square of twice the weights,
as well as both the difference
$\left( R_{i_j^k} - S_{i_j^k} \right)$ and the adjacent difference
$\left( R_{i_{j+1}^k} - S_{i_{j+1}^k} \right)$
(these two differences are independent, so the variance of their difference
is the sum of their variances).
Including both differences cancels any linear trend that might occur
when there is miscalibration, with the minor disadvantage
of decreasing the resolution of the estimate slightly (due to the averaging
across adjacent indices).
The estimator in~(\ref{estimator}) is a special case of that proposed
by~\cite{tygert_biostats}.

\section{ECEs and ICIs are popular yet poor metrics for calibration}
\label{ece_ici}

The ECE (``empirical calibration error,'' ``estimated calibration error,''
or ``expected calibration error'') and its other, less common name
(``ICI'' --- ``integrated calibration index'')
refer to longstanding metrics of calibration.
Statistical estimators for them are inherently problematic
when used for finite sample sizes (that is, for any finite data set), however.
Such estimators involve binning into histograms
or use of kernel density estimation.
The present paper uses the Kuiper metric of~(\ref{kuiper1}) and~(\ref{kuiper2})
instead, for the following reasons.

\cite{arrieta-ibarra-gujral-tannen-tygert-xu}
proved rigorously together with illustrative practical examples
that the ECE or ICI require ``observations whose density is infinitely higher
than what the ECCEs [Kuiper metrics] require to attain the same consistency
and power against any fixed alternative distribution,
where `infinitely higher' means asymptotically, in the limit of the sample size
tending to infinity (or in the limit of the statistical confidence level
tending to 100\%)\dots.
The ECEs also exhibit an extreme dependence on the choice of bins,
with different choices of bins yielding significantly different values
for the ECE metrics; choosing among the possible binnings can be confusing,
yet makes all the difference. In contrast, the ECCEs [Kuiper metrics] yield
trustworthy results without needing such large numbers of observations
and without needing to set any parameters.''

\cite{lee-huang-hassani-dobriban} proved rigorously
that this failure of the ECE or ICI to meet even the most basic criteria
for statistical acceptability (non-trivial power and/or consistency
asymptotically) is unavoidable in any method based on histograms
or kernel density estimation.
Lists of references which point out problems with the ECE or ICI are available
from~\cite{tygert_full} and~\cite{arrieta-ibarra-gujral-tannen-tygert-xu},
among others.
The abstract of the latter concludes, ``metrics based on binning
or kernel density estimation unavoidably must trade-off statistical confidence
for the ability to resolve variations as a function
of the predicted probability or vice versa.
Widening the bins or kernels averages away random noise
while giving up some resolving power. Narrowing the bins or kernels
enhances resolving power while not averaging away as much noise.
The cumulative [Kuiper] methods do not impose such an explicit trade-off.
Considering these results, practitioners probably
should adopt the cumulative approach as a standard for best practices.''
The present paper follows the recommended best practices.

\section{Boosting for multi-calibration}
\label{boosting}

This appendix reviews and amends an algorithm introduced
by~\cite{hebert-johnson-kim-reingold-rothblum} to improve calibration
over many subpopulations simultaneously.

The original algorithm improved calibration by regressing the residuals
(residual compared to perfect calibration) of successive improvements
to calibration, repeatedly cycling through all subpopulations targeted.
The original algorithm's improvements to calibration involved binning
the scores and adjusting the predictions in each bin
to make them perfectly calibrated on average.

The present paper considers a closely related technique,
which enables much higher efficiency.
We consider as input a model for classification (or regression)
that has been trained to yield predictions for half the provided training set.
Then, given any model for classification (or regression),
we fit the new model to the other half of the provided training set,
using not only the original covariates but also the predicted confidences
from the already trained input model.
That is, we fit the new model using the original covariates
augmented with the predicted probabilities from the already trained model.
Moreover, we can repeat this fitting several times
(using the same half of the training data set),
each time updating the scores used to augment the original variates
with the confidences for the new predictions.

\section{A synthetic example with closed-form analytic expressions}
\label{synthetic}

This appendix synthesizes a data set for which closed-form analytic expressions
are available for the Kuiper metrics, associated standard deviations,
and metrics of multi-calibration.

We choose an odd positive integer $q$ and set the number of observations
in the full population to $n_0 = q(q+1)$.

We set the scores to be linear functions of the index, namely,
\begin{equation}
\label{scorex}
S_j = \frac{2j + q}{2(q+1)^2}
\end{equation}
for $j = 1$, $2$, \dots, $n_0$.

We use uniform weights,
\begin{equation}
W_j = 1
\end{equation}
for $j = 1$, $2$, \dots, $n_0$.

We form $q$ blocks of $q+1$ responses;
for $j = 1$, $2$, \dots, $q$:
\begin{equation}
\label{responsex1}
R_i = 1
\end{equation}
for $i = (j-1)(q+1) + 1$, $(j-1)(q+1) + 2$, \dots, $(j-1)(q+1) + j$,
and
\begin{equation}
\label{responsex0}
R_i = 0
\end{equation}
for $i = (j-1)(q+1) + j + 1$, $(j-1)(q+1) + j + 2$, \dots, $(j-1)(q+1) + q+1$.

Combining~(\ref{responsex1}) and~(\ref{responsex0}) yields that
\begin{equation}
\sum_{i = (j-1)(q+1) + 1}^{j(q+1)} R_i = j
\end{equation}
for $j = 1$, $2$, \dots, $q$.
It follows from~(\ref{scorex}) that
\begin{equation}
\sum_{i = (j-1)(q+1) + 1}^{j(q+1)} S_i = j
\end{equation}
for $j = 1$, $2$, \dots, $q$, too.

The cumulative sum of the responses is therefore always greater than
or equal to the cumulative sum of the scores, for every index.
The maximum deviation will hence occur at one of the indices
$(j-1)(q+1) + j$ for $j = 1$, $2$, \dots, $q$.
The cumulative response at such an index is
\begin{equation}
\label{cumresponse}
\sum_{i=1}^{(j-1)(q+1) + j} R_i = \frac{j (j + 1)}{2}
\end{equation}
for $j = 1$, $2$, \dots, $q$.
The corresponding cumulative score is
\begin{equation}
\label{cumscore}
\sum_{i=1}^{(j-1)(q+1) + j} S_i
= \frac{j(q+2)(j(q+2)-q-1)}{2(q+1)^2}
\end{equation}
for $j = 1$, $2$, \dots, $q$.

The Kuiper metric depends on the maximum deviation
between the cumulative response and the cumulative score.
Combining~(\ref{cumresponse}) and~(\ref{cumscore}) yields
that the maximum deviation is
\begin{equation}
\label{maxdev}
\max_{1 \le j \le q} \frac{j[(j+1)(q+1)^2 - (q+2)(j(q+2)-q-1)]}{2(q+1)^2}.
\end{equation}
Differentiating the argument to the maximum in~(\ref{maxdev})
with respect to $j$ and setting to 0 yields that the argmax is $j = (q+1) / 2$,
so~(\ref{maxdev}) becomes
\begin{equation}
\label{maxdevsimp}
\max_{1 \le j \le q} \frac{j[(j+1)(q+1)^2 - (q+2)(j(q+2)-q-1)]}{2(q+1)^2}
= \frac{2q + 3}{8}.
\end{equation}
Normalizing~(\ref{maxdevsimp}) by the size ($n_0$) of the full population
yields the Kuiper metric
\begin{equation}
\label{kuiperfull}
D_0 = \frac{2q + 3}{8q(q+1)}.
\end{equation}

For subpopulations, we select the middlemost blocks of indices
\begin{equation}
i_j^k = k(q+1) + j
\end{equation}
for $j = 1$, $2$, \dots, $n_k$, with
\begin{equation}
n_k = n_0 - 2k(q+1) = (q-2k)(q+1)
\end{equation}
for $k = 1$, $2$, \dots, $\ell$, with $\ell = (q-1)/2$.

For every subpopulation, $k = 1$, $2$, \dots, $\ell$,
the maximum cumulative deviation again occurs at
the index $i_j^k = (q-1)(q+1)/2 + (q+1)/2 = q(q+1)/2$,
which determines the Kuiper metrics for the subpopulations.
The sum of the responses at that point is
\begin{equation}
\label{cumresponses}
\sum_{j=k(q+1)+1}^{q(q+1)/2} R_j = \sum_{j=k}^{(q+1)/2} j
= \frac{(q+1)(q+3)}{8} - \frac{k(k+1)}{2}
\end{equation}
and the sum of the scores is
\begin{equation}
\label{cumscores}
\sum_{j=k(q+1)+1}^{q(q+1)/2} S_j
= \frac{(q(q+1)/2 + k(q+1) + 1 + q)(q(q+1)/2 - k(q+1))}{2(q+1)^2}
= \frac{q(q+2) - 4k(k+1)}{8}
\end{equation}
for $k = 1$, $2$, \dots, $\ell$, with $\ell = (q-1)/2$.
Subtracting~(\ref{cumscores}) from~(\ref{cumresponses}) yields
the maximum cumulative deviation:
\begin{equation}
\label{maxcumdev}
\frac{2q + 3}{8}
\end{equation}
for $k = 1$, $2$, \dots, $\ell$, with $\ell = (q-1)/2$.
Normalizing~(\ref{maxcumdev}) by the size ($n_k$) of the subpopulation yields
the Kuiper metric
\begin{equation}
\label{kuiperex}
D_k = \frac{2q + 3}{8(q-2k)(q+1)}
\end{equation}
for $k = 1$, $2$, \dots, $\ell$, with $\ell = (q-1)/2$.

To calculate the standard deviation from~(\ref{stddev}),
we need to compute the sum of the scores and the sum
of the squares of the scores, for the full population
and every subpopulation considered.
The sum of the scores is
\begin{equation}
\label{notsquares}
\sum_{j = k(q+1) + 1}^{n_0 - k(q+1)} S_j
= \frac{(k(q+1) + 1 + q(q+1) - k(q+1) + q)(q-2k)(q+1)}{2(q+1)^2}
= \frac{(q-2k)(q+1)}{2}
\end{equation}
for subpopulation $k$, for $k = 0$, $1$, \dots, $\ell$.
The sum of the squares of the scores is
\begin{multline}
\label{squares}
\sum_{j = k(q+1)+1}^{n_0 - k(q+1)} (S_j)^2
= \frac{\sum_{j = k(q+1)+1}^{q(q+1) - k(q+1)} (4j^2 + 4jq + q^2)}{4(q+1)^4} \\
= \frac{2q(q(q+1) - k(q+1) + k(q+1)+1)(q-2k)(q+1) + q^2(q-2k)(q+1)
+ 4 \sum_{j = k(q+1)+1}^{q(q+1) - k(q+1)} j^2}{4(q+1)^4} \\
= \frac{1}{12(q+1)^4}
\Big( 4q^6 - 12kq^5 + 18q^5 + 12k^2 q^4 - 48kq^4 + 33q^4 - 8k^3 q^3
+ 36k^2 q^3 - 78kq^3 + 31q^3 \\ -\,24k^3 q^2 + 36k^2 q^2 - 66kq^2 + 14q^2
- 24k^3 q + 12k^2 q - 28kq + 2q - 8k^3 - 4k \Big)
\end{multline}
for subpopulation $k$, for $k = 0$, $1$, \dots, $\ell$.
Subtracting~(\ref{squares}) from~(\ref{notsquares}) yields
the standard deviation from~(\ref{stddev}):
\begin{multline}
\label{stddevex}
\frac{1}{n_k} \sqrt{\sum_{j = k(q+1) + 1}^{n_0 - k(q+1)} S_j (1 - S_j)} \\
= \frac{1}{(q-2k)(q+1)^3 \sqrt{12}} \sqrt{\phantom{\overline{q}}}
\overline{2q^6 + 12q^5 - 12k^2 q^4 - 12k q^4 + 27q^4
+ 8k^3 q^3 - 36k^2 q^3 - 42kq^3 + 29q^3\ } \hfill \\
\hfill \overline{\ +\,24k^3 q^2 - 36k^2 q^2 - 54k q^2 + 16q^2
+ 24k^3 q - 12k^2 q - 32kq + 4q + 8k^3 - 8k\ } \\
\end{multline}
for subpopulation $k$, for $k = 0$, $1$, \dots, $\ell$.

Combining~(\ref{kuiperex}) and~(\ref{stddevex}) yields that the metric $M$
of multi-calibration from~(\ref{multimetric}) is
\begin{multline}
\label{multiex}
M = \max_{0 \le k \le \ell}
\frac{(2q+3)\sqrt{2q^5 + 12q^4 + 27q^3 + 29q^2 + 16q + 4}}{8(q+1)}
\bigg/\sqrt{\phantom{\overline{q}}} \overline{2q^7 + 12q^6\ } \\
\overline{\ -\,12k^2 q^5 - 12k q^5 + 27q^5
+ 8k^3 q^4 - 36k^2 q^4 - 42kq^4 + 29q^4
+ 24k^3 q^3 - 36k^2 q^3 - 54k q^3 + 16q^3\ } \\
\overline{\ +\,24k^3 q^2 - 12k^2 q^2 - 32kq^2 + 4q^2 + 8k^3 q - 8kq\ }.
\end{multline}
The partial derivative (with respect to $k$) of the square
of the longest square-root in~(\ref{multiex}) is
\begin{equation}
-24kq^5 + 24k^2 q^4 - 12q^5 - 72kq^4 + 72k^2 q^3 - 42q^4 - 72kq^3 + 72k^2 q^2
- 54 q^3 - 24k q^2 + 24k^2 q - 32q^2 - 8q,
\end{equation}
which is negative for all $k$ in the interval $[0, \ell]$,
where $\ell = (q-1)/2$.
Thus, the maximum in~(\ref{multiex}) occurs at $k = \ell$
and so evaluating~(\ref{multiex}) at $k = \ell$ yields
\begin{equation}
\label{multimetricex}
M = \frac{2q+3}{8(q+1)}
\sqrt{\frac{2q^5 + 12q^4 + 27q^3 + 29q^2 + 16q + 4}
{3q^6 + 15q^5 + 29q^4 + 27q^3 + 13q^2 + 3q}}.
\end{equation}

The maximum of the Kuiper metrics $D_0$, $D_1$, \dots, $D_{\ell}$
from~(\ref{kuiperfull}) and~(\ref{kuiperex}) is
\begin{equation}
\label{nonorm}
\max_{0 \le k \le \ell} D_k = D_{\ell} = \frac{2q + 3}{8(q+1)},
\end{equation}
which is larger than $M$ in~(\ref{multimetricex}) by roughly $\sqrt{3q/2}$
when $q$ is sufficiently large. The following appendix refers
to~(\ref{nonorm}) as the ``multi-ablate'' metric.

\section{Additional numerical results}
\label{additional}

This final appendix reports further results on the data sets
from Subsections~\ref{acs2019} and~\ref{kddcup1998}.
Tables~\ref{acsTABLETLGBMClassifier}--\ref{acsSMARTPHONELogisticRegression}
pertain to the data set described in Subsection~\ref{acs2019}.
Tables~\ref{kddcupLGBMClassifier}--\ref{kddcupLogisticRegression}
pertain to the data set described in Subsection~\ref{kddcup1998}.

In the tables, the headings have the following meanings:
\begin{itemize}
\item ``error'' refers to the value obtained by subtracting the accuracy
from 1; that is, the error is the fraction of classifications
that are incorrect.
\item ``Kuiper'' refers to the metric of calibration (for the full population)
defined in~(\ref{kuiper1}) or, equivalently, in~(\ref{kuiper2}).
\item ``multi-cal.''\ refers to the metric of multi-calibration defined
in~(\ref{multimetric}).
\item ``multi-ablate'' refers to the same metric of multi-calibration defined
in~(\ref{multimetric}), but retaining only $D_k$ and no other factors,
thus reporting simply $\max_{0 \le k \le \ell} D_k$,
with no adjustment for the standard deviation given by~(\ref{stddev}).
\item ``expectation'' refers to the expected value
of the Kuiper metric $D_k$ calculated under the null hypothesis
of perfect calibration for a subpopulation (the $k$th)
which attains the maximum value of the ``multi-ablate'' metric
$\max_{0 \le k \le \ell} D_k$,
with the expected value being roughly 1.6 times the standard deviation
calculated via~(\ref{stddev}). For discussion of the factor of 1.6 used here,
see the paragraph following~(\ref{stddev}).
\item ``before calibration'' refers to the performance of the model
on the test set following only training on the training set.
\item ``Platt scaling'' refers to further calibrating
via cross-validation using Platt (temperature) scaling.
\item ``isotonic regression'' refers to further calibrating
via cross-validation using isotonic regression.
\item ``LightGBM'' refers to three repetitions of augmentation
as in Appendix~\ref{boosting}, using LightGBM (which performs boosting
internally to itself).
\item ``decision trees'' refers to three repetitions of augmentation
as in Appendix~\ref{boosting}, using vanilla (unboosted) decision trees.
\item ``logistic regression'' refers to three repetitions of augmentation
as in Appendix~\ref{boosting}, using logistic regression.
\end{itemize}
Subsection~\ref{models} provides full details on the models, ``LightGBM,''
``decision trees,'' and ``logistic regression.''
The lines with ``$\pm$'' in the tables report twice the standard error
of the mean for each metric on the previous line.
Thus, the intervals of values reported in the tables
are roughly 95\% confidence intervals for the true mean values.
The standard error is taken over three distinct random seeds
for generating the subpopulations via Algorithm~\ref{generation}.
The estimate of the standard error includes Bessel's correction
(so divides by 2 rather than 3 in the estimate for the variance
of the 3 observed values for the metric of multi-calibration,
which makes the estimate unbiased and more conservative).
The variation due to randomization in Algorithm~\ref{generation}
appears to be rather small in the experiments reported here.

\begin{table}[p]
\begin{center}
\begin{tabular}{llllllll}
&&& \makebox[0pt][l]{after calibration with} && \makebox[0pt][l]{after three augmentations} \\
&&& \makebox[0pt][l]{cross-validation via:}
&& \makebox[0pt][l]{via the specified method:} \\\hline
&& before & Platt & isotonic && decision & logistic \\
metric && calibration & scaling \, & regression\phantom{--} & LightGBM 
& trees 
& regression \\\hline
error && .2148 & .2273 & {\bf .2025} & .2033 & .2108 & .2650 \\
Kuiper && .04085 & .2599 & {\bf .01379} & .03886 & .04161 & .06345 \\
multi- && .07205 & .2997 & {\bf .03907} & .07085 & .06734 & .1296 \\
\phantom{m} cal. && \negphantom{$\pm$}$\pm$.00353 & \negphantom{$\pm$}$\pm$.00669 & \negphantom{$\pm$}$\pm${\bf .00336} & \negphantom{$\pm$}$\pm$.00495 & \negphantom{$\pm$}$\pm$.00031 & \negphantom{$\pm$}$\pm$.00000 \\\hline
multi- && .4189 & .4778 & .3765 & .3731 & .4605 & .5983 \\
ablate && \negphantom{$\pm$}$\pm$.02447 & \negphantom{$\pm$}$\pm$.00509 & \negphantom{$\pm$}$\pm$.00853 & \negphantom{$\pm$}$\pm$.04365 & \negphantom{$\pm$}$\pm$.12810 & \negphantom{$\pm$}$\pm$.05048 \\
expec- && .1965 & .2432 & .2092 & .1966 & .1966 & .1966 \\
tation && \negphantom{$\pm$}$\pm$.06411 & \negphantom{$\pm$}$\pm$.04683 & \negphantom{$\pm$}$\pm$.05236 & \negphantom{$\pm$}$\pm$.06102 & \negphantom{$\pm$}$\pm$.06102 & \negphantom{$\pm$}$\pm$.06102 \\\hline
\end{tabular}
\end{center}
\vspace{-1em}
\caption{LightGBM model on the test set of the U.S.\ Census Bureau's American Community Survey for tablet computers; the error when using no covariates at all is 0.2650.}
\label{acsTABLETLGBMClassifier}
\end{table}

\begin{table}[p]
\begin{center}
\begin{tabular}{llllllll}
&&& \makebox[0pt][l]{after calibration with} && \makebox[0pt][l]{after three augmentations} \\
&&& \makebox[0pt][l]{cross-validation via:}
&& \makebox[0pt][l]{via the specified method:} \\\hline
&& before & Platt & isotonic && decision & logistic \\
metric && calibration & scaling \, & regression\phantom{--} & LightGBM 
& trees 
& regression \\\hline
error && .2075 & .2294 & .2074 & {\bf .2033} & .2108 & .2650 \\
Kuiper && .02280 & .2560 & {\bf .01573} & .03886 & .04161 & .06345 \\
multi- && {\bf .05267} & .2949 & .05269 & .07085 & .06734 & .1296 \\
\phantom{m} cal. && \negphantom{$\pm$}$\pm${\bf .00325} & \negphantom{$\pm$}$\pm$.00398 & \negphantom{$\pm$}$\pm$.00301 & \negphantom{$\pm$}$\pm$.00495 & \negphantom{$\pm$}$\pm$.00031 & \negphantom{$\pm$}$\pm$.00000 \\\hline
multi- && .4534 & .4765 & .3912 & .3731 & .4605 & .5983 \\
ablate && \negphantom{$\pm$}$\pm$.05802 & \negphantom{$\pm$}$\pm$.00121 & \negphantom{$\pm$}$\pm$.03335 & \negphantom{$\pm$}$\pm$.04365 & \negphantom{$\pm$}$\pm$.12810 & \negphantom{$\pm$}$\pm$.05048 \\
expec- && .1965 & .2432 & .2092 & .1966 & .1966 & .2269 \\
tation && \negphantom{$\pm$}$\pm$.06411 & \negphantom{$\pm$}$\pm$.04683 & \negphantom{$\pm$}$\pm$.05236 & \negphantom{$\pm$}$\pm$.06102 & \negphantom{$\pm$}$\pm$.06102 & \negphantom{$\pm$}$\pm$.02159 \\\hline
\end{tabular}
\end{center}
\vspace{-1em}
\caption{Decision-tree model on the test set of the U.S.\ Census Bureau's American Community Survey for tablet computers; the error when using no covariates at all is 0.2650.}
\label{acsTABLETDecisionTreeClassifier}
\end{table}

\begin{table}[p]
\begin{center}
\begin{tabular}{llllllll}
&&& \makebox[0pt][l]{after calibration with} && \makebox[0pt][l]{after three augmentations} \\
&&& \makebox[0pt][l]{cross-validation via:}
&& \makebox[0pt][l]{via the specified method:} \\\hline
&& before & Platt & isotonic && decision & logistic \\
metric && calibration & scaling \, & regression\phantom{--} & LightGBM 
& trees 
& regression \\\hline
error && .2650 & .2463 & .2291 & {\bf .2033} & .2108 & .2650 \\
Kuiper && .06391 & .2345 & {\bf .02007} & .03886 & .04161 & .06345 \\
multi- && .1302 & .2792 & {\bf .06360} & .07085 & .06734 & .1296 \\
\phantom{m} cal. && \negphantom{$\pm$}$\pm$.00000 & \negphantom{$\pm$}$\pm$.00636 & \negphantom{$\pm$}$\pm${\bf .00972} & \negphantom{$\pm$}$\pm$.00495 & \negphantom{$\pm$}$\pm$.00031 & \negphantom{$\pm$}$\pm$.00000 \\\hline
multi- && .6002 & .4751 & .5039 & .3731 & .4605 & .5983 \\
ablate && \negphantom{$\pm$}$\pm$.05013 & \negphantom{$\pm$}$\pm$.00634 & \negphantom{$\pm$}$\pm$.08941 & \negphantom{$\pm$}$\pm$.04365 & \negphantom{$\pm$}$\pm$.12810 & \negphantom{$\pm$}$\pm$.05048 \\
expec- && .2266 & .2326 & .2218 & .1966 & .1966 & .2269 \\
tation && \negphantom{$\pm$}$\pm$.02159 & \negphantom{$\pm$}$\pm$.05165 & \negphantom{$\pm$}$\pm$.02515 & \negphantom{$\pm$}$\pm$.06102 & \negphantom{$\pm$}$\pm$.06102 & \negphantom{$\pm$}$\pm$.02159 \\\hline
\end{tabular}
\end{center}
\vspace{-1em}
\caption{Logistic-regression model on the test set of the U.S.\ Census Bureau's American Community Survey for tablet computers; the error when using no covariates at all is 0.2650.}
\label{acsTABLETLogisticRegression}
\end{table}

\begin{table}[p]
\begin{center}
\begin{tabular}{llllllll}
&&& \makebox[0pt][l]{after calibration with} && \makebox[0pt][l]{after three augmentations} \\
&&& \makebox[0pt][l]{cross-validation via:}
&& \makebox[0pt][l]{via the specified method:} \\\hline
&& before & Platt & isotonic && decision & logistic \\
metric && calibration & scaling \, & regression\phantom{--} & LightGBM 
& trees 
& regression \\\hline
error && .07863 & .08192 & {\bf .07331} & .08647 & .09740 & .1087 \\
Kuiper && .03530 & .3832 & {\bf .01273} & .02568 & .03478 & .09470 \\
multi- && .04683 & .4064 & {\bf .03213} & .04173 & .07664 & .1147 \\
\phantom{m} cal. && \negphantom{$\pm$}$\pm$.00198 & \negphantom{$\pm$}$\pm$.00960 & \negphantom{$\pm$}$\pm${\bf .00226} & \negphantom{$\pm$}$\pm$.00073 & \negphantom{$\pm$}$\pm$.03826 & \negphantom{$\pm$}$\pm$.00446 \\\hline
multi- && .4593 & .4715 & .4598 & .4849 & .5087 & .6074 \\
ablate && \negphantom{$\pm$}$\pm$.07141 & \negphantom{$\pm$}$\pm$.00188 & \negphantom{$\pm$}$\pm$.04422 & \negphantom{$\pm$}$\pm$.03613 & \negphantom{$\pm$}$\pm$.05003 & \negphantom{$\pm$}$\pm$.04958 \\
expec- && .1361 & .2629 & .1744 & .1087 & .1087 & .1087 \\
tation && \negphantom{$\pm$}$\pm$.08641 & \negphantom{$\pm$}$\pm$.04788 & \negphantom{$\pm$}$\pm$.03740 & \negphantom{$\pm$}$\pm$.07011 & \negphantom{$\pm$}$\pm$.07011 & \negphantom{$\pm$}$\pm$.07011 \\\hline
\end{tabular}
\end{center}
\vspace{-1em}
\caption{LightGBM model on the test set of the U.S.\ Census Bureau's American Community Survey for laptop computers; the error when using no covariates at all is 0.1087.}
\label{acsLAPTOPLGBMClassifier}
\end{table}

\begin{table}[p]
\begin{center}
\begin{tabular}{llllllll}
&&& \makebox[0pt][l]{after calibration with} && \makebox[0pt][l]{after three augmentations} \\
&&& \makebox[0pt][l]{cross-validation via:}
&& \makebox[0pt][l]{via the specified method:} \\\hline
&& before & Platt & isotonic && decision & logistic \\
metric && calibration & scaling \, & regression\phantom{--} & LightGBM 
& trees 
& regression \\\hline
error && {\bf .07901} & .1087 & .08031 & .08647 & .09740 & .1087 \\
Kuiper && {\bf .01415} & .3736 & .01609 & .02568 & .03478 & .09470 \\
multi- && .04531 & .4003 & {\bf .03536} & .04173 & .07664 & .1147 \\
\phantom{m} cal. && \negphantom{$\pm$}$\pm$.01833 & \negphantom{$\pm$}$\pm$.00842 & \negphantom{$\pm$}$\pm${\bf .00248} & \negphantom{$\pm$}$\pm$.00073 & \negphantom{$\pm$}$\pm$.03826 & \negphantom{$\pm$}$\pm$.00446 \\\hline
multi- && .4937 & .4660 & .4727 & .4849 & .5087 & .6074 \\
ablate && \negphantom{$\pm$}$\pm$.04733 & \negphantom{$\pm$}$\pm$.00170 & \negphantom{$\pm$}$\pm$.06111 & \negphantom{$\pm$}$\pm$.03613 & \negphantom{$\pm$}$\pm$.05003 & \negphantom{$\pm$}$\pm$.04958 \\
expec- && .1361 & .2629 & .1744 & .1087 & .1087 & .1818 \\
tation && \negphantom{$\pm$}$\pm$.08641 & \negphantom{$\pm$}$\pm$.04788 & \negphantom{$\pm$}$\pm$.03740 & \negphantom{$\pm$}$\pm$.07011 & \negphantom{$\pm$}$\pm$.07011 & \negphantom{$\pm$}$\pm$.03708 \\\hline
\end{tabular}
\end{center}
\vspace{-1em}
\caption{Decision-tree model on the test set of the U.S.\ Census Bureau's American Community Survey for laptop computers; the error when using no covariates at all is 0.1087.}
\label{acsLAPTOPDecisionTreeClassifier}
\end{table}

\begin{table}[p]
\begin{center}
\begin{tabular}{llllllll}
&&& \makebox[0pt][l]{after calibration with} && \makebox[0pt][l]{after three augmentations} \\
&&& \makebox[0pt][l]{cross-validation via:}
&& \makebox[0pt][l]{via the specified method:} \\\hline
&& before & Platt & isotonic && decision & logistic \\
metric && calibration & scaling \, & regression\phantom{--} & LightGBM 
& trees 
& regression \\\hline
error && .1087 & .1087 & .08798 & {\bf .08647} & .09740 & .1087 \\
Kuiper && .09312 & .3682 & {\bf .01039} & .02568 & .03478 & .09470 \\
multi- && .1131 & .3971 & {\bf .04089} & .04173 & .07664 & .1147 \\
\phantom{m} cal. && \negphantom{$\pm$}$\pm$.00452 & \negphantom{$\pm$}$\pm$.00907 & \negphantom{$\pm$}$\pm${\bf .00387} & \negphantom{$\pm$}$\pm$.00073 & \negphantom{$\pm$}$\pm$.03826 & \negphantom{$\pm$}$\pm$.00446 \\\hline
multi- && .6112 & .4676 & .5334 & .4849 & .5087 & .6074 \\
ablate && \negphantom{$\pm$}$\pm$.04863 & \negphantom{$\pm$}$\pm$.00115 & \negphantom{$\pm$}$\pm$.07509 & \negphantom{$\pm$}$\pm$.03613 & \negphantom{$\pm$}$\pm$.05003 & \negphantom{$\pm$}$\pm$.04958 \\
expec- && .1808 & .2430 & .1432 & .1087 & .1087 & .1818 \\
tation && \negphantom{$\pm$}$\pm$.03679 & \negphantom{$\pm$}$\pm$.01243 & \negphantom{$\pm$}$\pm$.02779 & \negphantom{$\pm$}$\pm$.07011 & \negphantom{$\pm$}$\pm$.07011 & \negphantom{$\pm$}$\pm$.03708 \\\hline
\end{tabular}
\end{center}
\vspace{-1em}
\caption{Logistic-regression model on the test set of the U.S.\ Census Bureau's American Community Survey for laptop computers; the error when using no covariates at all is 0.1087.}
\label{acsLAPTOPLogisticRegression}
\end{table}

\begin{table}[p]
\begin{center}
\begin{tabular}{llllllll}
&&& \makebox[0pt][l]{after calibration with} && \makebox[0pt][l]{after three augmentations} \\
&&& \makebox[0pt][l]{cross-validation via:}
&& \makebox[0pt][l]{via the specified method:} \\\hline
&& before & Platt & isotonic && decision & logistic \\
metric && calibration & scaling \, & regression\phantom{--} & LightGBM 
& trees 
& regression \\\hline
error && .06080 & .06254 & {\bf .04585} & .04970 & .06256 & .06254 \\
Kuiper && .02301 & .4079 & {\bf .01168} & .01192 & .02966 & .08586 \\
multi- && .03009 & .4259 & {\bf .01778} & .01903 & .04015 & .09898 \\
\phantom{m} cal. && \negphantom{$\pm$}$\pm$.00415 & \negphantom{$\pm$}$\pm$.00963 & \negphantom{$\pm$}$\pm${\bf .00262} & \negphantom{$\pm$}$\pm$.00038 & \negphantom{$\pm$}$\pm$.00522 & \negphantom{$\pm$}$\pm$.00665 \\\hline
multi- && .3539 & .4650 & .2345 & .2522 & .5262 & .4892 \\
ablate && \negphantom{$\pm$}$\pm$.02678 & \negphantom{$\pm$}$\pm$.00139 & \negphantom{$\pm$}$\pm$.00873 & \negphantom{$\pm$}$\pm$.03019 & \negphantom{$\pm$}$\pm$.14432 & \negphantom{$\pm$}$\pm$.13285 \\
expec- && .1824 & .2557 & .2335 & .1128 & .1128 & .1128 \\
tation && \negphantom{$\pm$}$\pm$.00703 & \negphantom{$\pm$}$\pm$.00588 & \negphantom{$\pm$}$\pm$.02589 & \negphantom{$\pm$}$\pm$.01410 & \negphantom{$\pm$}$\pm$.01410 & \negphantom{$\pm$}$\pm$.01410 \\\hline
\end{tabular}
\end{center}
\vspace{-1em}
\caption{LightGBM model on the test set of the U.S.\ Census Bureau's American Community Survey for smartphones; the error when using no covariates at all is 0.06254.}
\label{acsSMARTPHONELGBMClassifier}
\end{table}

\begin{table}[p]
\begin{center}
\begin{tabular}{llllllll}
&&& \makebox[0pt][l]{after calibration with} && \makebox[0pt][l]{after three augmentations} \\
&&& \makebox[0pt][l]{cross-validation via:}
&& \makebox[0pt][l]{via the specified method:} \\\hline
&& before & Platt & isotonic && decision & logistic \\
metric && calibration & scaling \, & regression\phantom{--} & LightGBM 
& trees 
& regression \\\hline
error && .05895 & .06254 & {\bf .04744} & .04970 & .06256 & .06254 \\
Kuiper && {\bf .008902} & .4068 & .01770 & .01192 & .02966 & .08586 \\
multi- && .01918 & .4236 & .02125 & {\bf .01903} & .04015 & .09898 \\
\phantom{m} cal. && \negphantom{$\pm$}$\pm$.00405 & \negphantom{$\pm$}$\pm$.00921 & \negphantom{$\pm$}$\pm$.00044 & \negphantom{$\pm$}$\pm${\bf .00038} & \negphantom{$\pm$}$\pm$.00522 & \negphantom{$\pm$}$\pm$.00665 \\\hline
multi- && .3083 & .4654 & .3103 & .2522 & .5262 & .4892 \\
ablate && \negphantom{$\pm$}$\pm$.04641 & \negphantom{$\pm$}$\pm$.00116 & \negphantom{$\pm$}$\pm$.07236 & \negphantom{$\pm$}$\pm$.03019 & \negphantom{$\pm$}$\pm$.14432 & \negphantom{$\pm$}$\pm$.13285 \\
expec- && .1824 & .2557 & .2335 & .1128 & .1128 & .1692 \\
tation && \negphantom{$\pm$}$\pm$.00703 & \negphantom{$\pm$}$\pm$.00588 & \negphantom{$\pm$}$\pm$.02589 & \negphantom{$\pm$}$\pm$.01410 & \negphantom{$\pm$}$\pm$.01410 & \negphantom{$\pm$}$\pm$.02174 \\\hline
\end{tabular}
\end{center}
\vspace{-1em}
\caption{Decision-tree model on the test set of the U.S.\ Census Bureau's American Community Survey for smartphones; the error when using no covariates at all is 0.06254.}
\label{acsSMARTPHONEDecisionTreeClassifier}
\end{table}

\begin{table}[p]
\begin{center}
\begin{tabular}{llllllll}
&&& \makebox[0pt][l]{after calibration with} && \makebox[0pt][l]{after three augmentations} \\
&&& \makebox[0pt][l]{cross-validation via:}
&& \makebox[0pt][l]{via the specified method:} \\\hline
&& before & Platt & isotonic && decision & logistic \\
metric && calibration & scaling \, & regression\phantom{--} & LightGBM 
& trees 
& regression \\\hline
error && .06254 & .06254 & .06768 & {\bf .05714} & .06256 & .06254 \\
Kuiper && .09031 & .3987 & {\bf .006544} & .01161 & .02966 & .08586 \\
multi- && .1032 & .4147 & .04495 & {\bf .02328} & .04015 & .09898 \\
\phantom{m} cal. && \negphantom{$\pm$}$\pm$.00657 & \negphantom{$\pm$}$\pm$.00855 & \negphantom{$\pm$}$\pm$.00000 & \negphantom{$\pm$}$\pm${\bf .00297} & \negphantom{$\pm$}$\pm$.00522 & \negphantom{$\pm$}$\pm$.00665 \\\hline
multi- && .4838 & .4631 & .3015 & .3187 & .5262 & .4892 \\
ablate && \negphantom{$\pm$}$\pm$.13255 & \negphantom{$\pm$}$\pm$.00040 & \negphantom{$\pm$}$\pm$.03733 & \negphantom{$\pm$}$\pm$.02098 & \negphantom{$\pm$}$\pm$.14432 & \negphantom{$\pm$}$\pm$.13285 \\
expec- && .1715 & .1923 & .09129 & .1128 & .1128 & .1692 \\
tation && \negphantom{$\pm$}$\pm$.02193 & \negphantom{$\pm$}$\pm$.04695 & \negphantom{$\pm$}$\pm$.01025 & \negphantom{$\pm$}$\pm$.01410 & \negphantom{$\pm$}$\pm$.01410 & \negphantom{$\pm$}$\pm$.02174 \\\hline
\end{tabular}
\end{center}
\vspace{-1em}
\caption{Logistic-regression model on the test set of the U.S.\ Census Bureau's American Community Survey for smartphones; the error when using no covariates at all is 0.06254.}
\label{acsSMARTPHONELogisticRegression}
\end{table}

\begin{table}[p]
\begin{center}
\begin{tabular}{llllllll}
&&& \makebox[0pt][l]{after calibration with} && \makebox[0pt][l]{after three augmentations} \\
&&& \makebox[0pt][l]{cross-validation via:}
&& \makebox[0pt][l]{via the specified method:} \\\hline
&& before & Platt & isotonic && decision & logistic \\
metric && calibration & scaling \, & regression\phantom{--} & LightGBM 
& trees 
& regression \\\hline
error && .3424 & .3393 & {\bf .3388} & .3453 & .3459 & .4688 \\
Kuiper && .08917 & .1538 & {\bf .006591} & .06773 & .008098 & .02284 \\
multi- && .08960 & .1539 & {\bf .01508} & .06840 & .01805 & .03667 \\
\phantom{m} cal. && \negphantom{$\pm$}$\pm$.00000 & \negphantom{$\pm$}$\pm$.00004 & \negphantom{$\pm$}$\pm${\bf .00019} & \negphantom{$\pm$}$\pm$.00010 & \negphantom{$\pm$}$\pm$.00161 & \negphantom{$\pm$}$\pm$.00972 \\\hline
multi- && .3935 & .4447 & .4185 & .3967 & .4073 & .4253 \\
ablate && \negphantom{$\pm$}$\pm$.02546 & \negphantom{$\pm$}$\pm$.04623 & \negphantom{$\pm$}$\pm$.03513 & \negphantom{$\pm$}$\pm$.03265 & \negphantom{$\pm$}$\pm$.12692 & \negphantom{$\pm$}$\pm$.04707 \\
expec- && .2363 & .2319 & .2364 & .2290 & .2290 & .2290 \\
tation && \negphantom{$\pm$}$\pm$.00279 & \negphantom{$\pm$}$\pm$.02261 & \negphantom{$\pm$}$\pm$.00311 & \negphantom{$\pm$}$\pm$.00151 & \negphantom{$\pm$}$\pm$.00151 & \negphantom{$\pm$}$\pm$.00151 \\\hline
\end{tabular}
\end{center}
\vspace{-1em}
\caption{LightGBM model on the test set of the 1998 KDD Cup for star donors; the error when using no covariates at all is 0.4688.}
\label{kddcupLGBMClassifier}
\end{table}

\begin{table}[p]
\begin{center}
\begin{tabular}{llllllll}
&&& \makebox[0pt][l]{after calibration with} && \makebox[0pt][l]{after three augmentations} \\
&&& \makebox[0pt][l]{cross-validation via:}
&& \makebox[0pt][l]{via the specified method:} \\\hline
&& before & Platt & isotonic && decision & logistic \\
metric && calibration & scaling \, & regression\phantom{--} & LightGBM 
& trees 
& regression \\\hline
error && .3514 & .3496 & .3509 & {\bf .3405} & .3459 & .4688 \\
Kuiper && {\bf .004652} & .1434 & .005189 & .06780 & .008098 & .02284 \\
multi- && .01383 & .1436 & {\bf .01344} & .06827 & .01805 & .03667 \\
\phantom{m} cal. && \negphantom{$\pm$}$\pm$.00144 & \negphantom{$\pm$}$\pm$.00009 & \negphantom{$\pm$}$\pm${\bf .00105} & \negphantom{$\pm$}$\pm$.00028 & \negphantom{$\pm$}$\pm$.00161 & \negphantom{$\pm$}$\pm$.00972 \\\hline
multi- && .3897 & .4681 & .3851 & .3914 & .4073 & .4253 \\
ablate && \negphantom{$\pm$}$\pm$.04151 & \negphantom{$\pm$}$\pm$.04725 & \negphantom{$\pm$}$\pm$.04249 & \negphantom{$\pm$}$\pm$.00828 & \negphantom{$\pm$}$\pm$.12692 & \negphantom{$\pm$}$\pm$.04707 \\
expec- && .2363 & .2319 & .2364 & .2290 & .2290 & .2317 \\
tation && \negphantom{$\pm$}$\pm$.00279 & \negphantom{$\pm$}$\pm$.02261 & \negphantom{$\pm$}$\pm$.00311 & \negphantom{$\pm$}$\pm$.00151 & \negphantom{$\pm$}$\pm$.00151 & \negphantom{$\pm$}$\pm$.02247 \\\hline
\end{tabular}
\end{center}
\vspace{-1em}
\caption{Decision-tree model on the test set of the 1998 KDD Cup for star donors; the error when using no covariates at all is 0.4688.}
\label{kddcupDecisionTreeClassifier}
\end{table}

\begin{table}[p]
\begin{center}
\begin{tabular}{llllllll}
&&& \makebox[0pt][l]{after calibration with} && \makebox[0pt][l]{after three augmentations} \\
&&& \makebox[0pt][l]{cross-validation via:}
&& \makebox[0pt][l]{via the specified method:} \\\hline
&& before & Platt & isotonic && decision & logistic \\
metric && calibration & scaling \, & regression\phantom{--} & LightGBM 
& trees 
& regression \\\hline
error && .4688 & .4607 & .4554 & {\bf .3405} & .3459 & .4688 \\
Kuiper && .02175 & .03736 & .009734 & .06780 & {\bf .008098} & .02284 \\
multi- && .03610 & .04623 & .02118 & .06827 & {\bf .01805} & .03667 \\
\phantom{m} cal. && \negphantom{$\pm$}$\pm$.01000 & \negphantom{$\pm$}$\pm$.00521 & \negphantom{$\pm$}$\pm$.00160 & \negphantom{$\pm$}$\pm$.00028 & \negphantom{$\pm$}$\pm${\bf .00161} & \negphantom{$\pm$}$\pm$.00972 \\\hline
multi- && .4255 & .4330 & .4037 & .3914 & .4073 & .4253 \\
ablate && \negphantom{$\pm$}$\pm$.04490 & \negphantom{$\pm$}$\pm$.06345 & \negphantom{$\pm$}$\pm$.02877 & \negphantom{$\pm$}$\pm$.00828 & \negphantom{$\pm$}$\pm$.12692 & \negphantom{$\pm$}$\pm$.04707 \\
expec- && .2317 & .2484 & .2328 & .2290 & .2290 & .2317 \\
tation && \negphantom{$\pm$}$\pm$.02245 & \negphantom{$\pm$}$\pm$.00783 & \negphantom{$\pm$}$\pm$.00749 & \negphantom{$\pm$}$\pm$.00151 & \negphantom{$\pm$}$\pm$.00151 & \negphantom{$\pm$}$\pm$.02247 \\\hline
\end{tabular}
\end{center}
\vspace{-1em}
\caption{Logistic-regression model on the test set of the 1998 KDD Cup for star donors; the error when using no covariates at all is 0.4688.}
\label{kddcupLogisticRegression}
\end{table}

\clearpage

\bibliography{paper}

\begin{thebibliography}{}

\bibitem[\protect\citename{Arrieta-Ibarra {\em et~al.},
  }2022]{arrieta-ibarra-gujral-tannen-tygert-xu}
Arrieta-Ibarra, Imanol, Gujral, Paman, Tannen, Jonathan, Tygert, Mark, \& Xu,
  Cherie. 2022.
\newblock Metrics of calibration for probabilistic predictions.
\newblock {\em J. Mach. Learn. Res.}, {\bf 23}, 1--54.

\bibitem[\protect\citename{Austin \& Steyerberg, }2019]{austin-steyerberg}
Austin, Peter, \& Steyerberg, Ewout. 2019.
\newblock The integrated calibration index (ICI) and related metrics for
  quantifying the calibration of logistic regression models.
\newblock {\em Stat. Med.}, {\bf 38}(21), 4051--4065.

\bibitem[\protect\citename{Br\"ocker, }2008]{brocker}
Br\"ocker, Jochen. 2008.
\newblock Some remarks on the reliability of categorical probability forecasts.
\newblock {\em Mon. Weather Rev.}, {\bf 136}(11), 4488--4502.

\bibitem[\protect\citename{Br\"ocker \& Smith, }2007]{brocker-smith}
Br\"ocker, Jochen, \& Smith, Leonard~A. 2007.
\newblock Increasing the reliability of reliability diagrams.
\newblock {\em Weather Forecast.}, {\bf 22}(3), 651--661.

\bibitem[\protect\citename{Gohar \& Cheng, }2023]{gohar-cheng}
Gohar, Usman, \& Cheng, Lu. 2023.
\newblock A survey on intersectional fairness in machine learning: notions,
  mitigation, and challenges.
\newblock {\em Pages  6619--6627 of:} Elkind, Edith (ed), {\em Proceedings of
  the Thirty-Second International Joint Conference on Artificial Intelligence}.
\newblock International Joint Conferences on Artificial Intelligence.

\bibitem[\protect\citename{Gopalan {\em et~al.}, }2023]{gopalan-kim-reingold}
Gopalan, Parikshit, Kim, Michael~P., \& Reingold, Omer. 2023.
\newblock Swap agnostic learning, or characterizing omniprediction via
  multicalibration.
\newblock {\em Pages  39936--39956 of:} Oh, Alice, Naumann, Tristan, Globerson,
  Amir, Saenko, Kate, Hardt, Moritz, \& Levine, Sergey (eds), {\em Proceedings
  of the 37th International Conference on Neural Information Processing
  Systems}.
\newblock Curran Associates.
\newblock No. 1736.

\bibitem[\protect\citename{Haghtalab {\em et~al.},
  }2023]{haghtalab-jordan-zhao}
Haghtalab, Nika, Jordan, Michael~I., \& Zhao, Eric. 2023.
\newblock A unifying perspective on multicalibration: game dynamics for
  multi-objective learning.
\newblock {\em Pages  72464--72506 of:} Oh, Alice, Naumann, Tristan, Globerson,
  Amir, Saenko, Kate, Hardt, Moritz, \& Levine, Sergey (eds), {\em Proceedings
  of the 37th International Conference on Neural Information Processing
  Systems}.
\newblock Curran Associates.
\newblock No. 3169.

\bibitem[\protect\citename{Hansen {\em et~al.},
  }2024]{hansen-devic-nakkiran-sharan}
Hansen, Dutch, Devic, Siddartha, Nakkiran, Preetum, \& Sharan, Vatsal. 2024.
\newblock When is multicalibration post-processing necessary?
\newblock {\em Pages  38383--38455 of:} Globerson, Amir, Mackey, Lester,
  Belgrave, Danielle, Fan, Angela, Paquet, Ulrich, Tomczak, Jakub, \& Zhang,
  Chiyuan (eds), {\em Proceedings of the 38th International Conference on
  Neural Information Processing Systems}.
\newblock Curran Associates.
\newblock No. 1213.

\bibitem[\protect\citename{Hebert-Johnson {\em et~al.},
  }2018]{hebert-johnson-kim-reingold-rothblum}
Hebert-Johnson, Ursula, Kim, Michael, Reingold, Omer, \& Rothblum, Guy. 2018.
\newblock Multicalibration: calibration for the (computationally identifiable)
  masses.
\newblock {\em Pages  1939--1948 of:} Dy, Jennifer, \& Krause, Andreas (eds),
  {\em Proceedings of the 35th International Conference on Machine Learning
  (ICML)}.
\newblock Proceedings of Machine Learning Research, vol. 80.
\newblock Microtome Press.

\bibitem[\protect\citename{Ke {\em et~al.},
  }2017]{ke-meng-finley-wang-chen-ma-ye-liu}
Ke, Guolin, Meng, Qi, Finley, Thomas, Wang, Taifeng, Chen, Wei, Ma, Weidong,
  Ye, Qiwei, \& Liu, Tie-Yan. 2017.
\newblock {L}ight{GBM}: a highly efficient gradient boosting decision tree.
\newblock {\em Pages  3149--3157 of:} {\em Proceedings of the 31st
  International Conference on Neural Information Processing Systems}.
\newblock Curran Associates.

\bibitem[\protect\citename{Kim {\em et~al.}, }2019]{kim-ghorbani-zou}
Kim, Michael~P., Ghorbani, Amirata, \& Zou, James. 2019.
\newblock Multiaccuracy: black-box post-processing for fairness in
  classification.
\newblock {\em Pages  247--254 of:} Conitzer, Vincent, Hadfield, Gillian, \&
  Vallor, Shannon (eds), {\em Proceedings of the 2019 AAAI/ACM Conference on
  Artificial Intelligence, Ethics, and Society (AIES)}.
\newblock Association for Computing Machinery.

\bibitem[\protect\citename{Kim {\em et~al.},
  }2022]{kim-kern-goldwasser-kreuter-reingold}
Kim, Michael~P., Kern, Christoph, Goldwasser, Shafi, Kreuter, Frauke, \&
  Reingold, Omer. 2022.
\newblock Universal adaptability: target-independent inference that competes
  with propensity scoring.
\newblock {\em Proc. Natl. Acad. Sci. U.S.A.}, {\bf 119}(4), 1--6.
\newblock No. e2108097119.

\bibitem[\protect\citename{Kolmogorov, }1933]{kolmogorov}
Kolmogorov, Andrei~N. 1933.
\newblock Sulla determinazione empirica di una legge di distribuzione ({O}n the
  empirical determination of a distribution function).
\newblock {\em Giorn. Ist. Ital. Attuar.}, {\bf 4}, 83--91.

\bibitem[\protect\citename{Kuiper, }1962]{kuiper}
Kuiper, Nicolaas~H. 1962.
\newblock Tests concerning random points on a circle.
\newblock {\em Proc. Koninklijke Nederlandse Akademie van Wetenschappen Series
  A}, {\bf 63}, 38--47.

\bibitem[\protect\citename{{La C}ava {\em et~al.}, }2023]{la-cava-lett-wan}
{La C}ava, William~G., Lett, Elle, \& Wan, Guangya. 2023.
\newblock Fair admission risk prediction with proportional multicalibration.
\newblock {\em Pages  350--378 of:} Mortazavi, Bobak~J., Sarker, Tasmie, Beam,
  Andrew, \& Ho, Joyce~C. (eds), {\em Proceedings of the Conference on Health,
  Inference, and Learning (CHIL)}.
\newblock Proceedings of Machine Learning Research, vol. 209.
\newblock Microtome Press.

\bibitem[\protect\citename{Lee {\em et~al.}, }2023]{lee-huang-hassani-dobriban}
Lee, Donghwan, Huang, Xinmeng, Hassani, Hamed, \& Dobriban, Edgar. 2023.
\newblock {T-C}al: an optimal test for the calibration of predictive models.
\newblock {\em J. Mach. Learn. Res.}, {\bf 24}, 1--72.

\bibitem[\protect\citename{Pedregosa {\em et~al.}, }2011]{scikit-learn}
Pedregosa, Fabian, Varoquaux, Ga\"e, Gramfort, Alexandre, Michel, Vincent,
  Thirion, Bertrand, Grisel, Olivier, Blondel, Mathieu, Prettenhofer, Peter,
  Weiss, Ron, Dubourg, Vincent, Vander{P}las, Jake, Passos, Alexandre,
  Cournapeau, David, Brucher, Matthieu, Perrot, Matthieu, \& Duchesnay,
  Edouard. 2011.
\newblock Scikit-learn: machine learning in {P}ython.
\newblock {\em Journal of Machine Learning Research}, {\bf 12}, 2825--2830.

\bibitem[\protect\citename{Pfisterer {\em et~al.},
  }2021]{pfisterer-kern-dandl-sun-kim-bischl}
Pfisterer, Florian, Kern, Christoph, Dandl, Susanne, Sun, Matthew, Kim,
  Michael~P., \& Bischl, Bernd. 2021.
\newblock mcboost: multi-calibration boosting for {R}.
\newblock {\em Journal of Open Source Software}, {\bf 6}(64), 1--3.
\newblock No. 3453.

\bibitem[\protect\citename{Shabat {\em et~al.}, }2020]{shabat-cohen-mansour}
Shabat, Eliran, Cohen, Lee, \& Mansour, Yishay. 2020.
\newblock Sample complexity of uniform convergence for multicalibration.
\newblock {\em Pages  13331--13340 of:} Larochelle, Hugo, Ranzato,
  Marc-Aurelio, Hadsell, Raia, Balcan, Maria-Florina, \& Lin, Hsuan-Tien (eds),
  {\em Proceedings of the 34th International Conference on Neural Information
  Processing Systems}.
\newblock Curran Associates.
\newblock No. 1118.

\bibitem[\protect\citename{Smirnov, }1939]{smirnov}
Smirnov, Nikolai. 1939.
\newblock On the estimation of the discrepancy between empirical curves of
  distribution for two independent samples.
\newblock {\em Bulletin Math\'ematique de l'Universit\'e de Moscou}, {\bf
  2}(2), 3--11.

\bibitem[\protect\citename{Tygert, }2021]{tygert_full}
Tygert, Mark. 2021.
\newblock Cumulative deviation of a subpopulation from the full population.
\newblock {\em J. Big Data}, {\bf 8}(117), 1--60.
\newblock Available at \url{https://arxiv.org/abs/2008.01779}.

\bibitem[\protect\citename{Tygert, }2023]{tygert_pvals}
Tygert, Mark. 2023.
\newblock Calibration of {P}-values for calibration and for deviation of a
  subpopulation from the full population.
\newblock {\em Adv. Comput. Math.}, {\bf 49}(70), 1--22.
\newblock Available at \url{https://arxiv.org/abs/2202.00100}.

\bibitem[\protect\citename{Tygert, }2024]{tygert_biostats}
Tygert, Mark. 2024.
\newblock {\em Cumulative differences between subpopulations versus body mass
  index in the Behavioral Risk Factor Surveillance System data}.
\newblock Tech. rept. 2411.07399. arXiv.
\newblock Available at \url{https://arxiv.org/abs/2411.07399}.

\bibitem[\protect\citename{Wilks, }2019]{wilks}
Wilks, Daniel~S. 2019.
\newblock {\em Statistical Methods in the Atmospheric Sciences}. 4th edn.
\newblock Elsevier.

\bibitem[\protect\citename{Wu {\em et~al.}, }2024]{wu-liu-cui-wu}
Wu, Jiayun, Liu, Jiashuo, Cui, Peng, \& Wu, Zhiwei~Steven. 2024.
\newblock Bridging multicalibration and out-of-distribution generalization
  beyond covariate shift.
\newblock {\em Pages  73036--73078 of:} Globerson, Amir, Mackey, Lester,
  Belgrave, Danielle, Fan, Angela, Paquet, Ulrich, Tomczak, Jakub, \& Zhang,
  Chiyuan (eds), {\em Proceedings of the 38th International Conference on
  Neural Information Processing Systems}.
\newblock Curran Associates.
\newblock No. 2325.

\bibitem[\protect\citename{Ye \& Li, }2024]{ye-li}
Ye, Hanxuan, \& Li, Hongzhe. 2024.
\newblock {\em Multicalibration for modeling censored survival data with
  universal adaptability}.
\newblock Tech. rept. 2405.15948. arXiv.
\newblock Available at \url{https://arxiv.org/abs/2405.15948}.

\end{thebibliography}
\bibliographystyle{authordate1}

\end{document}